\documentclass{article}
\usepackage{spconf,graphicx}
\usepackage[fleqn]{amsmath}
\usepackage{amsfonts}
\usepackage{booktabs}
\usepackage{diagbox}
\usepackage{makecell}
\usepackage{siunitx}
\usepackage{graphicx}
\usepackage{multirow}
\usepackage{hyperref}
\usepackage[bottom]{footmisc}
\usepackage{adjustbox}

\newcommand{\xv}{\boldsymbol{x}}
\newcommand{\axv}{\widetilde{\boldsymbol{x}}}
\newcommand{\zv}{\boldsymbol{z}}
\newcommand{\tv}{\boldsymbol{t}}
\newcommand{\cv}{\boldsymbol{c}}
\newcommand{\Eb}{\mathbb{E}}

\newcommand{\KL}{D_{\mathrm{KL}}}
\newcommand*{\dif}{\mathop{}\!\mathrm{d}}

\newcommand{\numberthis}{\refstepcounter{equation}\tag{\theequation}}

\ninept

\title{MAXIMIZING MUTUAL INFORMATION FOR TACOTRON}
\name{Peng Liu$^\dagger$, Xixin Wu$^\ddagger$, Shiyin Kang$^\dagger$, 
Guangzhi Li$^\dagger$, Dan Su$^\dagger$, Dong Yu$^\dagger$}
\address{
  $^\dagger$ Tencent AI Lab \\
  $^\ddagger$ Department of Systems Engineering and Engineering Management,\\
       The Chinese University of Hong Kong, China\\
 \texttt{\{feanorliu, shiyinkang, guangzhilei, dansu, dyu\}@tencent.com} \\
 \texttt{wuxx@se.cuhk.edu.hk}}
\begin{document}
%
\maketitle
\begin{abstract}
  End-to-end speech synthesis methods already achieve
  close-to-human quality performance. However compared to
  HMM-based and NN-based frame-to-frame regression methods, 
  they are prone to some synthesis errors, such as missing or repeating 
  words and incomplete synthesis.
  We attribute the comparatively high utterance
  error rate to the local information preference of conditional
  autoregressive models, and the ill-posed training objective of the model,
  which describes mostly the training status of the autoregressive module, 
  but rarely that of the condition module.
  Inspired by InfoGAN, we propose to
  maximize the mutual information between the text condition and the
  predicted acoustic features to strengthen the dependency between them for CAR
  speech synthesis model, which would alleviate the local information
  preference issue and reduce the utterance error rate.
  The training objective of maximizing mutual information can be considered as a metric
  of the dependency between the autoregressive module and the condition module.
  Experiment results show that our method can reduce the utterance error rate
  \footnote{Code is available at \url{https://github.com/bfs18/tacotron2}}.
\end{abstract}
\section{Introduction}
\label{sec:intro}

Tacotron \cite{tacotron} and Tacotron2 \cite{tacotron2} are 
conditional autoregressive (CAR) models trained with 
teacher forcing \cite{teacher-forcing}. 
The condition is summarized from the input text with attention 
mechanism \cite{bahdanau-attention}.
Transformer-TTS \cite{transformer-tts} can be considered 
as another instance of CAR model, with effective utilization of
self-attention mechanism \cite{self-attention}. 
Such architecture can be trained in an end-to-end way, so it has a much shorter pipeline and
needs less expert knowledge and human labor.
It is flexible enough to adapt for speaking style \cite{prosody, style-tok}
and multi-speaker \cite{deep-voice2, multi-skpr}.
In addition, 
it is easy to be combined with neural vocoder \cite{wavernn, wavenet, pwavenet}
to enhance the synthesized waveform quality.

Training with teacher forcing induces a mismatch between the training period 
and the inference period, usually known as exposure bias \cite{exposure-bias}.
Even worse, it strengthens the 
local information preference \cite{vlae} for the CAR model. We explain the 
local information preference intuitively first.
At each time step during training, the CAR
model receives
a teacher forcing input and a conditional input. 
The teacher forcing input is one previous time 
step from the target. The conditional input is the text to be synthesized.
If the CAR model learns to copy the teacher forcing input, 
or to predict the target totally depending on teacher forcing input 
without using the conditional information,
it still gets small training root mean square error (RMSE). 
Finally the model, which achieves small RMSE, may not learn to depend on
the condition at all. So at the inference period, the CAR model generates results that
have nothing to do with the condition.
Note that local information preference still exists even if teacher 
forcing is not used. When a random variable $\xv$ admits autoregressive dependency
over a conditional random variable $\zv$, 
i.e. $p(\xv|\zv)=\prod_{i} (\xv_i|\xv_{<i},\zv)$, an universal function
approximator, such as RNNs used in the CAR model, can in theory represent the distribution without 
condition on $\zv$ \cite{vlae}. 

The local information preference weakens the 
dependency between the predicted acoustic features and the text condition 
when training a CAR speech synthesis model.
In most cases, the CAR speech synthesis model learns to depend on the 
text condition to predict the acoustic features. However they are prone to bad cases. 
We argue that this is caused by the local information preference of the model.
Since the model prefers predicting the acoustic features from the teacher forcing input
at training stage, 
it does not model the dependency between the text condition and the predicted acoustic features 
sufficiently. If we can strengthen the dependency, we may reduce the bad-case rate.

In \cite{infogan}, the authors propose a information-theoretic regularization for 
generative adversarial networks (GAN) \cite{gan} to learn a set of disentangled latent codes.
The authors separate GAN's input noise vector into incompressible noise and 
latent codes with factorized distribution. But the generator of 
GAN is free to ignore the additional 
latent codes and predicts observations only conditioning on the incompressible noise.
To eliminate such trivial solutions, the authors maximize the mutual information between
the latent codes and the observations for GAN. This leads to the InfoGAN model. 
The idea is straightforward. Since the the mutual
dependency between two variables can be measured by mutual information,
maximizing mutual information (MMI) would strengthen the dependency between 
the laten codes and the observations,
and hence eliminate the trivial solutions that
the GAN's generator models the observations without depending 
on the latent codes.
Inspired by InfoGAN, we propose to maximize the mutual information between the text
condition and the predicted acoustic features
to strengthen the dependency for CAR speech synthesis
models. This would alleviate the local information preference problem and reduce the
rate of bad cases. 

Viewing from another perspective, 
the reconstruction error only reflects the training status
of the autoregressive module since we can get small reconstruction error even if the 
model generates random human-voice-like acoustic features. The mutual information objective
can remedy the weakness of the original training objective.

In the following, we begin with explaining 
the local information preference formally for CAR 
speech synthesis models and review the existing designs in Tacotron which prevent
the model from predicting the target totally depending on teacher forcing input
in section \ref{sec:car}.
Then we explain our method and provide the experiment results in section \ref{sec:mmi} and 
\ref{sec:exp}.
Finally, we conclude our work in section \ref{sec:con}.

\section{Why CAR TTS model tends to ignore the text condition}
\label{sec:car}
  In this section we first explain local information preference for CAR model formally. 
  Then we explain why Tacotron still works though it tends to ignore the text condition.

\subsection{Variational encoder-decoder perspective of CAR model}
  Usually we perform maximum likelihood estimation (MLE) to train a CAR speech synthesis model. 
  And the model communicates information form the text to the acoustic features 
  through the time-aligned latent variables.
  Such latent variables exist in various speech recognition
  and synthesis systems, such as the hidden states in the HMM-based speech synthesis 
  system, the forward-backward search matrix in the CTC recognizer, and the attention variables
  in Tacotron. We can formalize the CAR speech synthesis model as a variational encoder-decoder
  (VED) \cite{ved}.
  We use $\tv$ and $\xv$ to represent a text and its corresponding acoustic features 
  in the training set.
  Since it is a CAR model, the conditional likelihood can be written as 
  $\log p_\theta(\xv|\tv)= \sum_{i=1}^{N} \log p_\theta(\xv_i|\xv_{<i}, \tv)$, 
  where $N$ is the number of acoustic frames in $\xv$.
  For simplicity, we suppose the distribution of the time-aligned
  latent variables, $\cv$, is factorizable,
  i.e. $\log p_\theta(\cv|\xv, \tv) = \sum_{i=1}^{N} \log p_\theta(\cv_i | \xv_{<i}, \tv )$.
  At least this is true for the ad hoc treatment of the 
  attention variables in Tacotron \cite{post-att}.
  Then $\log p_\theta(\xv|\tv) = \log \int_{\cv} p_\theta(\xv, \cv | \tv) \dif \cv = \sum_{i=1}^{N} \log \int_{\cv_i} p(\xv_i, \cv_i |\xv_{<i}, \tv) \dif \cv_i$.
  The training objective is to maximize the sum of the conditional likelihood of
  each $\tv$, $\xv$ pair in the training set. For a training pair at time
  step $i$:
  \[ 
        \log p_\theta(\xv_i|\xv_{<i}, \tv) = \KL(q_\phi(\cv_i|\xv_{<i}, \tv) || p_\theta(\cv_i | \xv_{\leq i}, \tv)) + \mathcal{L}(\theta, \phi, \xv, \tv) \numberthis \label{eq:loss_0}\\
  \]
  The first RHS term is the KL divergence of the encoder approximation from the model posterior (
  it is the posterior because it has access to the current $\xv_i$). 
  The second RHS term is the variational lower bound. 
  Since the $\mathrm{KL}$-divergence term is always non-negative:
  \begin{align*}
      & \log p_\theta(\xv_i|\xv_{<i}, \tv) \geq \mathcal{L}(\theta, \phi, \xv, \tv) \\
    = & \Eb_{q_\phi(\cv_i | \xv_{<i}, \tv)}[-\log q_\phi (\cv_i | \xv_{<i}, \tv) + \log p_\theta (\xv_i, \cv_i | \xv_{<i}, \tv)] \\
    = & -\KL(q_\phi(\cv_i | \xv_{<i}, \tv) || p_\theta(\cv_i | \xv_{<i}, \tv)) \\
      & + \Eb_{q_\phi(\cv_i | \xv_{<i}, t)} [ \log p_\theta (\xv_i | \xv_{<i}, \cv_i)] \numberthis \label{eq:loss_1} \\
    = & \log p_\theta(\xv_i | \xv_{<i}, \tv) - \KL(q_\phi(\cv_i|\xv_{<i}, \tv)||p_\theta(\cv_i|\xv_{\leq i}, \tv)) \numberthis \label{eq:loss_2}     
  \end{align*}
  In the above equations, 
  $\phi$ is the attention encoder parameters and $\theta$ is the autoregressive decoder 
  parameters for a CAR speech synthesis model.
  Since we suppose the text communicates information to the acoustic features only through
  the time-aligned latent variables, by Bayes rule, 
  we have $p(\xv_i, \cv_i | \xv_{<i}, \tv) = p(\xv_i | \xv_{<i}, \cv_{i})p(\cv_{i} | \xv_{<i}, \tv)$
  used in Eq \ref{eq:loss_1}.
  Note that the variational encoder-decoder formalization is a bit different from the original VAE. 
  From Eq \ref{eq:loss_0}, we can see that $q_\phi(\cv_i | \xv_{<i}, \tv)$ is used to approximate
  the model posterior distribution, $p_\theta(\cv_i | \xv_{\leq i}, \tv)$.
  However, it does not use information from the current $\xv_i$, 
  because $\xv_i$ is the acoustic feature frame to predict at inference time step $i$,
  we cannot use it as the input to the encoder.
  We can use $q_\phi(\cv_i | \xv_{<i}, \tv)$ as the prior 
  distribution, $p_\theta(\cv_i | \xv_{<i}, \tv)$,
  then the KL-divergence term becomes 0 in Eq \ref{eq:loss_1}. 
  If we use a deterministic function to calculate $\cv_i$, 
  Eq \ref{eq:loss_1} becomes the training objective of Tacotron.
  In Eq \ref{eq:loss_2}, the $\mathrm{KL}$-divergence term is 0
  only when $\xv_i$ and $\cv_i$ are conditionally independent.
  In such case, the time-aligned latent variables, $\cv$, are meaningless.
  If the model learns meaningful time-aligned latent variables, the 
  $\mathrm{KL}$-divergence term is positive.
  When trained to maximize $\log p_\theta(\xv_{i} | \xv_{<i}, \tv)$, 
  the model would not learn meaningful time-aligned latent variables to avoid 
  the extra cost if $\xv_{<i}$ contains enough information to predict $\xv_i$.
  Since the time aligned latent variables are the bridges that communicate
  information from text to acoustic features, 
  the model cannot exploit the text efficiently without the latent variables.
  we arrive at a similar conclusion as 
  variational lossy autoencoder (VLAE):
  information can be modeled locally by the CAR model will be modeled locally
  without using information from the time-aligned latent variables, 
  only the remainder will be modeled using them \cite{vlae}.
  We argue that this is one of 
  the possible reasons why attention mechanism cannot learn alignment
  under some bad configurations \cite{xixin-att}.

\subsection{Why Tacotron learns to condition on text}
  We argue that Tacotron learns to condition on the text mainly because of several designs:
  the reduction window, the large frame shift and the dropout in the decoder prenet.
  Reduction window is a frame dropout mechanism like the word dropout used in 
  VAE language model (VAELM) \cite{optimization-challenge}
  to weaken the connection between autoregressive steps. 
  Setting reduction factor to 5 \cite{tacotron}
  can be considered as dropping 80\% frames at equal intervals. This is a bit different from
  dropping words randomly to a certain percentage in VAELM, but they work in a similar way.

  We can use the Euclidean distance (ED) between the teacher forcing input and the acoustic target as a metric for information
  locality (ED relates to the RMSE used at training).
  If the ED is smaller, it is easier for the CAR model to predict the target only 
  based on the teacher forcing input without using text information. We list
  the frame averaged ED of mean-std normalized log mel wrapped
  short-time Fourier transform (STFT) magnitude for 
  different configurations for the LJSpeech dataset \cite{ljspeech} in Table \ref{tab:mel_mse}.
  If a reduction window is used, we repeat the teacher forcing input reduction factor times to 
  make the number of frames consistent with that of the target.
  From Table \ref{tab:mel_mse}, we can see that using larger reduction factor and frame shift
  could increase the ED between the teacher forcing input and the acoustic target, 
   which indicates that the connection between them is weakened.
  To achieve smaller training RMSE,
  the model has to depend more on the text.
  In \cite{gan-tts}, the authors point out that the decoder prenet dropout in
  Tacotron could make the model condition more on the input text. 
  Intuitively, the dropout makes the teacher forcing input incomplete,
  so the model has to condition more on the text to reconstruct the target.
  It is reported in \cite{tacotron2} that 
  significantly more pronunciation issues are observed when using 5ms frame shift.
  In \cite{s2s}, the authors report that a narrower prenet bottleneck is critical
  in picking up attention during training. Decreasing the prenet bottleneck
  size would compress the teacher forcing input and increase the information gap.
  In conclusion, increasing the information gap between the teacher forcing input 
  and target is vital for Tacotron to achieve acceptable performance.
  Also dropping teacher forcing input frames randomly to a certain percentage
  is a cheap trick to make the
  model more robust,
  which is not applied to Tacotron in previous works.

    \begin{table}
    \caption{Euclidean distance (ED) between the teacher forcing input and the acoustic target
             for different reduction factor and frame shift (frame length is 4 times of frame shift) 
             configurations for LJSpeech.}
    \label{tab:mel_mse}
    \centering
    \renewcommand{\arraystretch}{0.8}
    \begin{tabular}{ccc}
        \toprule
                    & \multicolumn{2}{c}{frame shift (ms)} \\ \cmidrule{2-3}
        reduction factor   & 5     & 12.5 \\
        \midrule
        1     & 0.289  & 0.374     \\
        \midrule
        2     & 0.367  & 0.507     \\
        \midrule
        5     & 0.516  & 0.751     \\
        \bottomrule
    \end{tabular}
    \end{table}

\section{Maximizing mutual information (MMI) for Tacotron}
\label{sec:mmi}
  Although the previous mentioned designs in Tacotron alleviate the local information
  preference, they weaken the autoregressive decoder and decrease the model's performance.
  A model using reduction factor 2 generates better perceptual results than one
  using reduction factor 5 \cite{tacotron}. This indicates that the more the 
  autoregressive model is weakened, the more drop in performance is induced.
  Even worse, Tacotron make mistakes, such as repeating words, omitting words and 
  incomplete sentences, which seldomly appear in HMM-based methods \cite{hmm-tts-1}
  or NN-based frame-to-frame regression methods 
  \cite{dnn-tts-3, dnn-tts-2, dnn-tts-1}. 
  The dependency between the predicted acoustic features
  and the text input in Tacotron is not sufficiently modeled. If the dependency is 
  sufficiently modeled and the model is penalized heavily when it makes mistakes 
  during training,
  the generated acoustic features should strictly follow the text. So we take the 
  InfoGAN approach that maximize the mutual information between
  the predicted acoustic features
  and the input text during training to strengthen the dependency between them.
\subsection{MMI with an auxiliary recognizer}
  The mutual information between the input text, $\tv$, and the
  predicted acoustic features, ${\axv}$, is
  \begin{alignat*}{2}
    &&&I(\axv; \tv) \\
        &=&&H(\tv) - H(\tv | \axv) \\
        &=&&\Eb_{\axv \sim p_\alpha(\xv)} [\Eb_{\tv \sim p_\alpha(\tv|\axv)} [\log p_\alpha(\tv|\axv)] ] + H(\tv)  \\
        &=&&\Eb_{\axv \sim p_\alpha(\xv)} [{D_{\mathrm{KL}}} (p_\alpha(\tv|\axv)||q_\beta(\tv|\axv)) + \Eb_{\tv \sim p_\alpha(\tv|\axv)} [\log q_\beta(\tv|\axv)]] \\
        &&&+ H(\tv) \numberthis \label{eq:mi_1} \\
        &\geq&&\Eb_{\axv \sim p_\alpha(\xv)} [ \Eb_{\tv \sim p_\alpha(\tv|\axv)} [\log q_\beta(\tv|\axv)] ] + H(\tv) \\
        &=&&\Eb_{\tv \sim p(\tv), \axv \sim p_\alpha(\xv|\tv)} [\log q_\beta(\tv|\axv)] + H(\tv) \numberthis \label{eq:mi_2}
  \end{alignat*}
  $\alpha=\{\theta, \phi\}$ is the CAR model parameters.
  In Eq \ref{eq:mi_1} we introduce an auxiliary distribution $q_\beta(\tv|\axv)$ to 
  approximate the posterior $p_\alpha(\tv|\axv)$ since it is intractable.
  The lower bound derivation uses the variational information maximization technique
  \cite{vim, infogan}.
  $H(\tv)$ is a constant for our 
  problem. From Eq \ref{eq:mi_2}, we can see that maximizing the mutual information
  between the input text and 
  the predicted acoustic features is equivalent to training an auxiliary recognizer which 
  maximizes the probability of recognizing the input 
  text from the predicted acoustic features with respect to the CAR model parameters, $\alpha$, 
  and the auxiliary recognizer parameters, $\beta$.
  This is intuitively sound. If the predicted acoustic features are consistently recognized as
  the input text, of course the model gets the correct result.
  Adding the mutual information term to the training objective in Eq \ref{eq:loss_0} 
  can penalize the model if it ignores the dependency 
  between the predicted acoustic features and the text.
  When this penalty is stronger than the $\mathrm{KL}$-divergence term 
  in Eq \ref{eq:loss_2},
  the model learns meaningful time-aligned latent variables to exploit the text.

\subsection{CTC recognizer for Tacotron}
  To keep the end-to-end property, we use a simple CTC recognizer as the auxiliary recognizer.
  The CTC recognizer uses the same convolution stack + bidirectional LSTM \cite{lstm} layer 
  structure as the Tacotron2's text encoder for simplicity
  except that the former has an extra CTC loss layer.
  Lack of a language model is usually considered as 
  a drawback of the CTC recognizer \cite{rnn-transducer}.
  However, this quite meets our demand, 
  since we do not want a language model to remedy the detected errors.
  Minimizing the CTC loss could strengthen the dependency between the predicted acoustic
  features and the input text during training. 

  The final loss function is:
  \begin{align*}
    \mathcal{L} = & |\xv_{mel} - \widetilde{\xv}_{mel}| + |\xv_{linear} - \widetilde{\xv}_{linear}| \\
    &+ CELoss(\xv_{stop}, \widetilde{\xv}_{stop}) + \lambda CTCLoss(\tv, \widetilde{\xv}_{mel}) \numberthis \label{eq:floss}
  \end{align*}
  $|\cdot|$ is L1 norm. 
  The first 2 RHS terms are the reconstruction losses for Mel spectrum and linear
  spectrum. Also the model minimize
  the cross entropy loss for stop tokens and the CTC loss between
  the predicted Mel spectrum  and the text to be synthesized.
  $\lambda$ controls the relative weight for 
  the CTC loss. 
  The linear loss is used, because we use the Griffin-Lim 
  algorithm to reconstruct waveforms to monitor the training progress.  

\section{Experiments}
\label{sec:exp}
  We show that maximizing the mutual information between the predicted acoustics 
  and the text to be synthesized can reduce the rate of bad case.

\subsection{Experiment setup}
  We use LJSpeech for English and 
  Databaker Chinese Standard Mandarin Speech Corpus 
  (db-CSMSC)\footnote{\url{https://www.data-baker.com/open_source.html}}
  for Mandarin Chinese in our experiments.
  LJSpeech contains \num[group-separator={,}]{13100} audio clips 
  of a single female speaker.
  We process the transcriptions with 
  Festival \cite{festival}
  to get the phoneme sequences.
  db-CSMSC contains \num[group-separator={,}]{10000} standard Mandarin
  sentences recorded by a single female native speaker and recorded in a professional
  recording studio. The dataset contains the Chinese character and pinyin
  transcriptions and hand-crafted time intervals. In our experiments, we
  only use the pinyin transcription and transfer the pinyin sequence 
  to a pinyin scheme which contains initials and sub-finals.
  Our pinyin scheme contains 
  much less units than the initial-final pinyin scheme. 
  It can alleviate
  the out-of-vocabulary and data sparsity problems.
  
  All the waveforms are downsampled to 16k Hz in our experiments.
  We extract 2048-point STFT magnitudes with Hanning window and wrap the 
  features with Mel filter to 80-band Mel spectrum.
  We use 12.5ms/50ms window shift for our experiments.
  Then a $\log$ operation is applied to linear spectrum and Mel spectrum.
  We use repeat padding for the training samples of different 
  lengths in a batch
  since zero padding would affect the batch normalization
  statistics.
  We use the Adam \cite{adam} optimizer with $\beta_1=0.9$, $\beta_2=0.999$ and
  $\epsilon=10^{-8}$. The initial
  learning rate is $0.002$ and starts to decay by a factor of $\sqrt{4000/step}$
  from $4000$ step \cite{t2t}.
  The gradient is clipped to maximum global norm of $1.0$ \cite{grad_clip}.
  We use Tacotron2 for our experiments. 
  For each dataset, we keep out 
  $1/64$ of the dataset as the validation set.
  For English, the test cases are randomly chosen from the 1132
  CMU\_ARCTIC \cite{cmu_arctic}
  sentences. For Mandarin Chinese, the test cases are chosen 
  from text of different domains.
  We use a test set of 100 sentences for listening test.
  The average numbers of words/characters and phonemes 
  in one utterance is 8.8 and 32.1 for English and 
  15.6 and 41.4 for Chinese.
  We use an open-sourced 
  WaveRNN vocoder \footnote{\url{https://github.com/fatchord/WaveRNN}}
  to reconstruct waveforms from Mel spectrums.

\subsection{UER and MOS for Tacotron-MMI}
In Tacotron2, the attention context is concatenated to the LSTM output and projected by 
a linear transform to predict the Mel spectrum. This means the predicated Mel
spectrum contains linear components of the text information. If we use this Mel spectrum
as the input to the CTC recognizer, the text information is too easily accessible for the
recognizer. This may cause the text information to be encoded in a pathological way in the Mel
spectrum and lead to a strict diagonal alignment map (one 
acoustic frame output for one phoneme input)
combined with location-sensitive attention. So before the linear transform operation,
we add an 
extra LSTM layer to mix the text information and acoustic information.
$\lambda$ is set to $1.0$ in our experiments and the checkpoint for evaluation
is selected at 200k training steps.

    \begin{table}
    \caption{Utterance error rate (UER) for different configurations.
    (RF is short for reduction factor and DFR is short for drop frame rate)}
    \label{tab:exp2_res}
    \centering
    \renewcommand{\arraystretch}{0.8}
    \begin{tabular}{ccccc}
    \toprule
    & & \multicolumn{2}{c}{RF 2} & RF 5 \\
    \cmidrule{3-5}
     corpus &  & DFR 0.0 & DFR 0.2 & DFR 0.0 \\
    \midrule
    \multirow{2}{*}{LJSpeech} & no MMI & 16\% & 15\% & 10\% \\
    \cmidrule{2-5}
                              & MMI & 10\% & 5\% & - \\         
    \midrule
    \multirow{2}{*}{db-CSMSC} & no MMI & 17\% & 12\% & 7\% \\
    \cmidrule{2-5}
                              & MMI & 5\% & 4\% & -\\

    \bottomrule
    \end{tabular}
    \end{table}

    \begin{table} 
    \caption{Mean opinion score (MOS) with 95\% confidence intervals for different configurations.}
    \label{tab:mos_res}
    \centering
    \renewcommand{\arraystretch}{0.8}
    \begin{tabular}{ccccc}
    \toprule
        & DFR 0.0 & DFR 0.2 & MMI + DFR 0.0 & MMI + DFR 0.2 \\
    \midrule
     MOS   & 3.84$\pm$0.16 &  3.92$\pm$0.17  &  3.83$\pm$0.14  & 3.87$\pm$0.15 \\
    \bottomrule
    \end{tabular}
    \end{table}

\begin{figure}[htb]
\begin{minipage}[b]{.48\linewidth}
  \centering
  \centerline{\includegraphics[width=4.0cm]{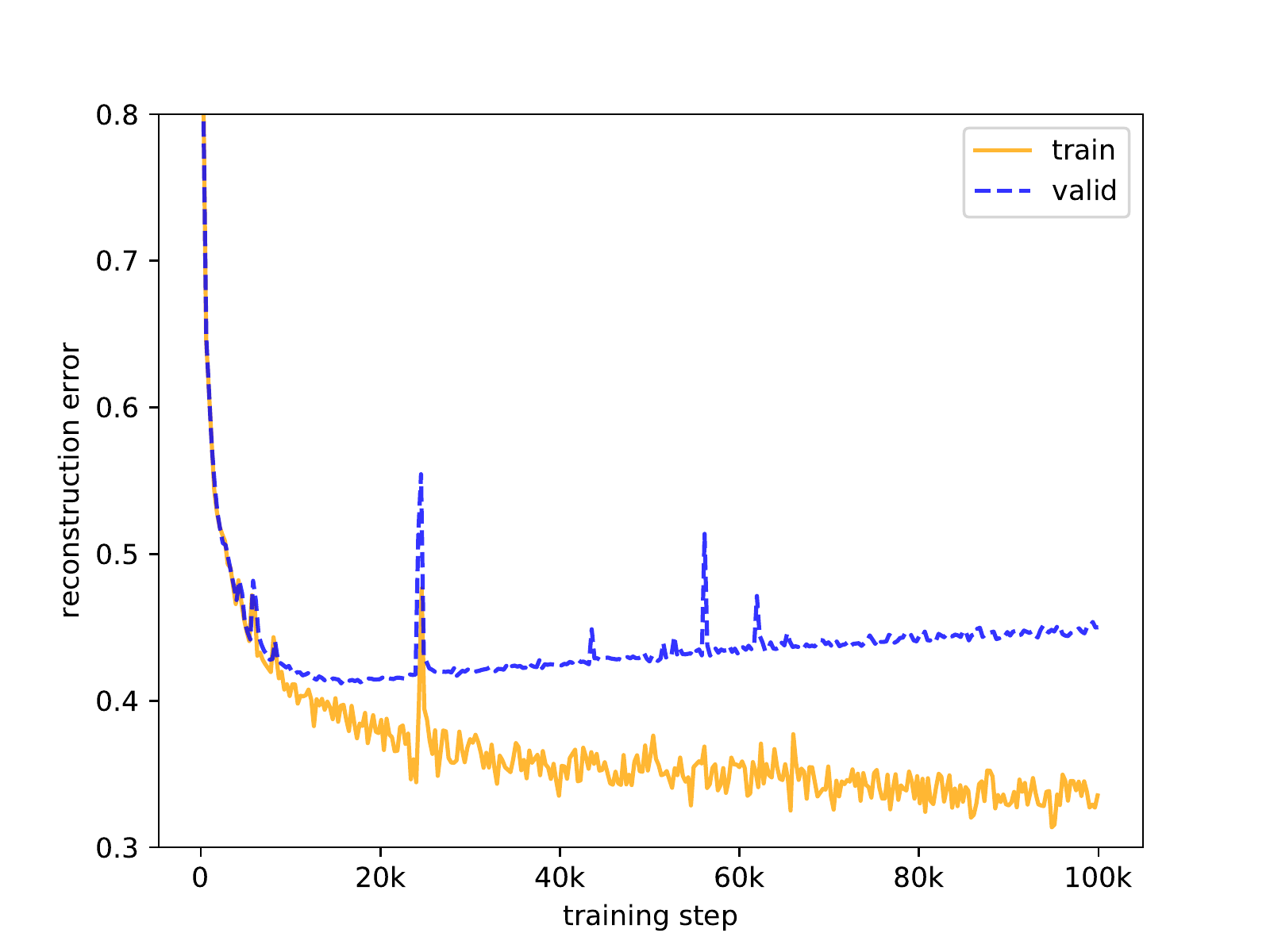}}
  \centerline{(a) Loss curve without MMI}\medskip
\end{minipage}
\hfill
\begin{minipage}[b]{0.48\linewidth}
  \centering
  \centerline{\includegraphics[width=4.0cm]{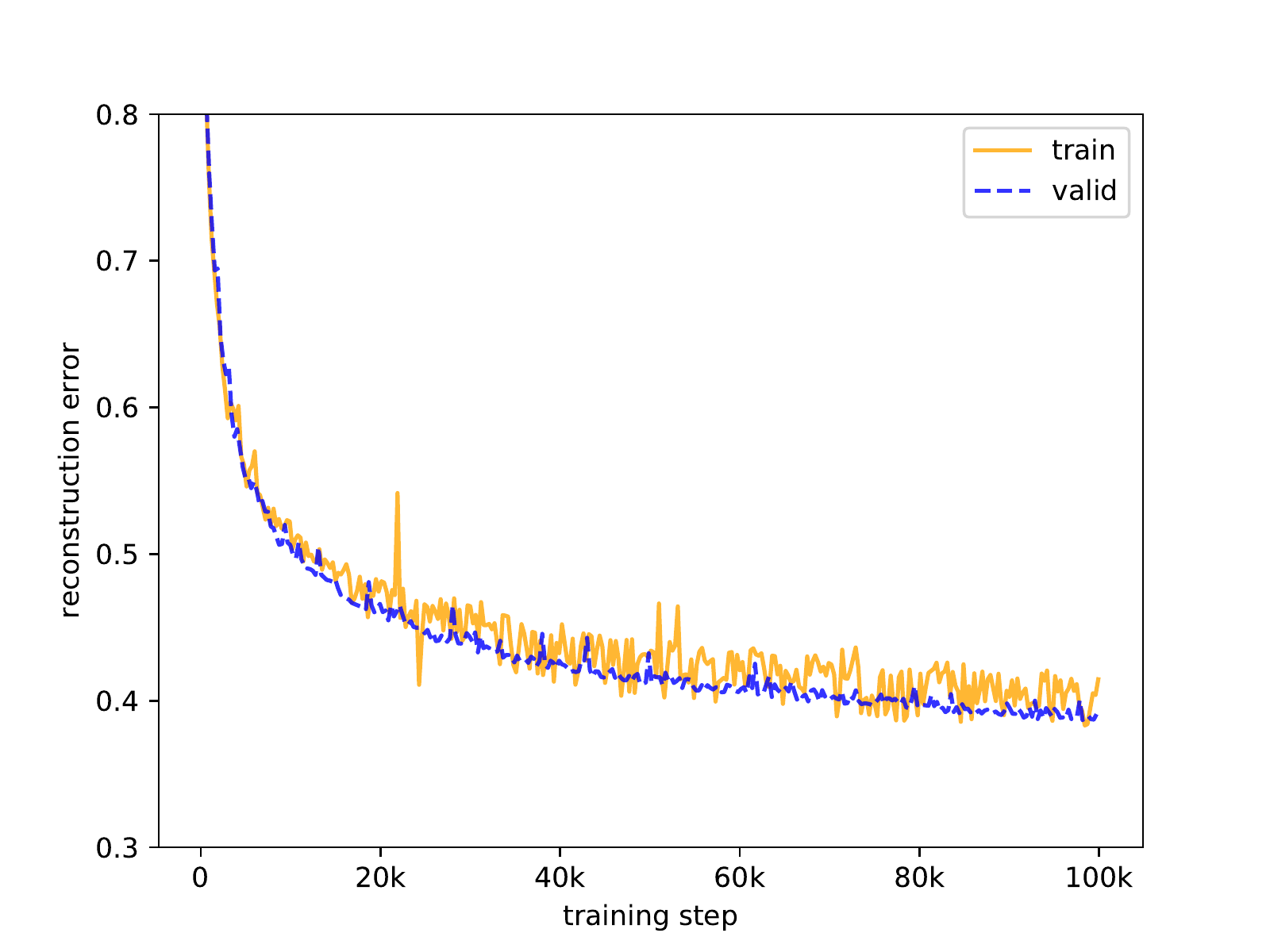}}
  \centerline{(b)  Loss curve with MMI}\medskip
\end{minipage}
\caption{Loss curves for training and validation set. 
 The x- and y-axis are training step and
 reconstruction error. Training curves are orange solid lines and 
 validation curves are blue dashed lines.
 (a) is the plot for a model trained without MMI and (b) is for one with MMI.}
\label{fig:loss}
\end{figure}

We use the default configuration with reduction factor (RF) 5 as the baseline
and we test the effectiveness of MMI and drop teacher forcing frame trick with RF 2.
The results are recorded in Table \ref{tab:exp2_res}.
We can see that both MMI and drop frame rate (DFR) 0.2 can reduce the 
utterance error rate (UER).
We observe that the gap of the reconstruction error
(the first 2 term in RHS of Eq. \ref{eq:floss})
between training and validation sets begins to increase from 10k steps
when MMI is not used.
This does not happen when MMI is used as depicted in Figure \ref{fig:loss}, 
indicating the MMI training objective prevents
the autoregressive module from fitting the non-linguistic detail
in acoustic features, while the original training objective is not a good
indicator as it does not take into consideration the dependency
between the acoustic features and text.

We conduct a mean opinion score (MOS) test to see whether 
the extra MMI objective would degenerate the synthesized
waveform quality or not. 
Only correctly synthesized waveforms are selected for this test.
From Table \ref{tab:mos_res}, 
we can see that Tacotron2 with DFR 0.2 achieves the best perceptual result
and the model with MMI achieves similar perceptual performance. 

\subsection{Discussion}
\label{sec:related}
  Many previous works focus on improving Tacotron's reliability.
  In \cite{gan-tts}, professor forcing is adopted to mitigate 
  the exposure bias induced by training with teacher forcing.
  The authors use diagonal attention penalty to enforce that 
  the alignment between the acoustic features and the text is 
  approximately diagonal in \cite{dig-att}.
  In \cite{att-loss}, the authors propose to use the alignment 
  information form hand-crafted labels or from an HHM-based system to guide the attention
  for Tacotorn. 
  Since a large body of legacy corpus and HMM-based systems exist, 
  this is an efficient way to improve Tacotron. 
  However, it is not trained in an end-to-end way.
  The implicit duration model of Tacotron uses alignment 
  information that is not self-contained.
  Transformer-TTS adopts self-attention structure to improve the training and inference 
  efficiency and to shorten the long range dependency path 
  between any two inputs at different
  time steps \cite{transformer-tts}. 

  In the speech-to-speech translation task \cite{s2s}, experiment results demonstrated 
  that the multi-task recognition loss worked, but without proper explanation.
  It can be explained by Eq. \ref{eq:mi_2}, where minimizing
  the multi-task recognition loss can be interpreted as maximizing
  the mutual information between the learned hidden representation
  and the corresponding text in that task. When training with
  the multi-task recognition loss, the learned hidden representation encodes
  more linguistic information rather than acoustic
  information only, results in a better fit for the speech translation task.

\section{Conclusion}
\label{sec:con}
In this paper we analyze why Tacotron is prone to synthesis errors.
In short, modeling the correlation between the text
and the acoustic features sufficiently is important to avoid
the bad cases.
To gain this objective, we propose to maximize the mutual information
between the text and the predicted acoustic features 
with an auxiliary CTC recognizer.
Experiment results show that our method can reduce the rate of
bad cases.
Besides our method can be trained in an end-to-end manner.
It keeps the short
pipeline of the original method.


\vfill\pagebreak

\label{sec:refs}
\bibliographystyle{IEEEbib}
\bibliography{refs}

\end{document}